\documentclass[12pt]{article}
\usepackage{amsthm,amsfonts,amsmath}
 \numberwithin{equation}{section}                             %
\newcommand{\version}{July 31, 2004}                          %
\font\notefont=cmsl8

\theoremstyle{plain}

\newtheorem{thm}{THEOREM}[section]
\newtheorem{lm}[thm]{LEMMA}

\theoremstyle{definition}

\theoremstyle{remark}
\newcommand{\upchi}{\raise1pt\hbox{$\chi$}}
\newcommand{\R}{{\mathord{\mathbb R}}}

\newcommand{\C}{{\mathord{\mathbb C}}}
\newcommand{\Z}{{\mathord{\mathcal  Z}}}

\newcommand{\tr}{{\mathord{\rm Tr}}}
\begin{document}

\title{\bf{The thermodynamic limit
  for matter interacting with Coulomb forces and with the quantized
  electromagnetic field:\\ 
  I. The lower bound}}
\author{\vspace{5pt} Elliott H. Lieb$^1$ and
Michael Loss$^{2}$ \\
\vspace{-4pt}\small{$1.$ Departments of Mathematics and Physics, Jadwin
Hall,} \\[-5pt]
\small{Princeton University, P.~O.~Box 708, Princeton, NJ
  08544}\\
\vspace{-4pt}\small{$2.$ School of Mathematics, Georgia Tech,
Atlanta, GA 30332}  \\ }
\date{\version}
\maketitle

\footnotetext
[1]{Work partially
supported by U.S. National Science Foundation
grant PHY 01-39984.}
\footnotetext
[2]{Work partially
supported by U.S. National Science Foundation
grant DMS 03-00349.\\
\copyright\, 2004 by the authors. This paper may be reproduced, in its
entirety, for non-commercial purposes.}

\begin{abstract} 
  
  The proof of the existence of the thermodynamic limit for electrons
  and nuclei interacting via the Coulomb potential, in the framework
  of non-relativistic quantum mechanics, was accomplished decades ago.
  This result did not take account of interactions caused by magnetic
  fields, however, (the spin-spin interaction, in particular) or of
  the quantized nature of the electromagnetic field.  Recent progress
  has made it possible to undertake such a proof in the context of
  non-relativistic QED.  This paper contains one part of such a proof
  by giving a lower bound to the free energy  which is
  proportional to the number of particles and which takes account of
  the fact that the field, unlike the particles, is never confined to
  a finite volume.  In the earlier proof the lower bound was a `two
  line' corollary of the `stability of matter'.  In QED the proof is
  much more complicated.

\end{abstract} %\hfill \version

\centerline{\bf } \medskip

\section{Introduction}

Some years ago the problem of proving the existence of the
thermodynamic limit for electrons, nuclei and other particles
interacting via Coulomb forces was settled in the context of the
non-relativistic Schr\"odinger equation \cite{LLeb}.  The key ingredients in
this proof, in broad outline, were:

a) The stability of matter of the second kind \cite{DL} (i.e., a lower
bound on the ground state energy proportional to the number of
particles), which led to an upper bound on the partition function
$\Z$, and hence a domain independent lower bound on $f$, the free
energy per particle.

b) A rigorous version of screening together with a variational argument
for a lower bound on $\Z$, which led to the fact that $f$ could only
decrease (with the density $\rho$ and inverse temperature $\beta = 1/k_B
T$ fixed) as the size of the domain $\Omega$ containing the particles
increases.  Charge neutrality is needed for this monotonicity of $f$ (but
not for the lower bound).  Since $f$ is bounded, this monotonicity
guarantees that $f$ has a limit as $|\Omega|$, the volume of $\Omega$,
tends to infinity.

Since then much progress has been made in understanding
non-relativistic quantum electrodynamics (QED) and it seems
appropriate now to try to extend the proof of the thermodynamic limit
to the QED case.  This is not just an idle exercise, for several new
matters of a physical nature, as well as a mathematical nature, arise.
Among these is the fact this model completely takes account of
everything that we know about low energy physics, except for the
hyperfine interaction (for which nuclear physics is necessary, as we
explain below), and except for the fact that the dynamics of the
particles (but not the electromagnetic field) is non-relativistic.
Indeed, no completely satisfactory relativistic Hamiltonian is
presently available and, therefore, the fully relativistic generalization
will have to await further developments.  Another problem, which is
yet to be resolved, is the renormalization of physical parameters in
order to deal with the infinities that arise as $\Lambda$, the
ultraviolet cutoff on the electromagnetic field, tends to infinity.

Otherwise, the theory is potentially complete, as we said, and an example
of this completeness is that it is not necessary to exclude the spin-spin
inter-electron magnetic interaction, as in \cite{LLeb}.  The usual non-QED
approximation is to mimic the interaction by a $r^{-3}$ spin-dependent
potential, which cannot possibly be stable, and which is, therefore,
omitted from discussion unless a hard core interaction is introduced to
stabilize it.  In contrast, in a full theory in which the magnetic field
$B(x)$ is a dynamical variable and the particles interact with the field
via a $\sigma \cdot B(x)$ term (but without any explicit spin-spin
interaction) is perfectly well behaved and stable and has all the right
physics in the classical limit.

(We note in passing that stability of matter requires more than just
the field energy to stabilize the $\sigma \cdot B(x)$ terms.  It also
requires the `kinetic' energy terms $(p+ eA(x)/c)^2$ to control the
$\sigma \cdot B(x)$ terms, and thereby stabilize the system.  In other
words, the terms $p\cdot A(x) + A(x)^2$ are essential for
understanding the interaction of particles with each other at small
distances; the dipole-dipole approximation while correct at large
distances, is certainly inadequate at short distances.)

Another major difference between the Schr\"odinger and the QED
theories of the thermodynamic limit is the necessity of treating the
thermodynamics of the field correctly.  In 1900 Planck \cite{Planck}
gave us the energy density of the pure electromagnetic field at
temperature $T$, which implies that the field cannot be confined to
the container $\Omega$ without invoking artificial constraints.  As we
shall explain in detail later, this requires us first to take a limit
in which the size of the universe $\mathcal{U}$ tends to infinity
(after subtracting the enormous pure Planck free energy) and
afterward to take the limit $|\Omega| \to \infty$.  Obviously, the
subtraction has to be done carefully and that is an exercise in
itself.

In this paper we consider topic a) above --- the upper bound on $\Z$ or
lower bound on $f$ (after taking the double limit, of course). We
shall reserve topic b) for later. In the previous work
\cite{LLeb} the upper bound required only a few lines, as we shall
explain below, but our QED setting presents significant difficulties
that have to be overcome. While the analog of the Dyson-Lenard lower
bound on the energy \cite{DL} is known for this QED case (see
\cite{LLS, BFG}), it is far from sufficient for obtaining the upper
bound on $\Z$.

\section{Basic Definitions }\label{sec2}

There are $N$ electrons with mass $m$ and charge $-e$.  These are
fermions with spin $1/2$.  There are also $K$ nuclei with several
kinds of masses $Mm$ (with $M>1800$ in nature), positive charges $Ze$
and statistics (Bose or Fermi) but, in order to simplify the notation,
we shall assume only one species with charge $Ze$ and mass $Mm$.  The
generalization to many species is trivial, the only significant point
being that that all the nuclei have a $Z$-value not greater than some
fixed number $Z$.  We also assume that the nuclei are point charges, the
generalization to smeared out nuclei being a trivial generalization.

The arena in which the particles reside is a large region $\Omega
\subset \R^3$, of unspecified shape for the present purposes, and
volume $|\Omega|$.  It is a subset of an even larger domain
$\mathcal{U}$, the `universe' which, for simplicity we take to be a
cube of side length $L$.  The boundary conditions of the EM field on
$\partial \mathcal{U}$ is, presumably, of no importance, so we take
periodic boundary conditions for simplicity (although it has to be
noted that changing the boundary conditions on the $\partial
\mathcal{U}$ changes the total energy (when the temperature is not
zero) by an amount far greater than the energy contained in $\Omega$).
One could dispense with the universe $\mathcal{U}$ by confining the EM
field to the box $\Omega$, but this would be questionable physically
and we shall not do so here. The two limits (i.e., with or without the
confinement of the field to $\Omega$) would be expected to yield the
same average energy density in the thermodynamic limit, but we prefer
to take nothing for granted.

The Hilbert space is 
\begin{equation}\label{hilbert} \mathcal{H} =
\mathcal{H}_{\mathrm{electron}} \otimes \mathcal{H}_{\mathrm{nuclei}} \otimes
\mathcal{F} \ , 
\end{equation} 
where $\mathcal{F}$ is the photon Fock space in $\mathcal{U}$ and
$\mathcal{H}_{\mathrm{electron}}$ is the antisymmetric tensor product
$\wedge_{i=1}^N L^2(\Omega; \C^2)$ appropriate for spin 1/2 fermions.
Likewise, $ \mathcal{H}_{\mathrm{nuclei}}$ is an antisymmetric tensor
product of $\wedge_{j=1}^K L^2(\Omega;\C^2)$ (for fermions) or a
symmetric tensor product of $K$ $L^2(\Omega)$ spaces (for bosons) or a
mixture of them in the case of several species.  A vector in
$\mathcal{H}$ is a function of $N$ electron coordinates and spins
$x_1, ...  , x_N; \sigma_1, ...  , \sigma_N$ and $K$ nuclear
coordinates (and possibly spins if they are fermions) $R_1, ...  ,
R_K$ with values in $\mathcal{F}$, i.e., it is a vector in
$\mathcal{F}$ that depends on the particle coordinates and spins.

\underline{\textit{Units}}: The physical units we shall employ here are 
$2mc^2$ for the energy and $\lambda_c /2$ for the length
(where $\lambda_c = \hbar/mc$ is the electron Compton wavelength).
The dimensionless fine structure constant is $\alpha = e^2/\hbar c$
(= 1/137 in nature). The electron charge is then $-\sqrt\alpha$ and the
nuclear charge is $Z\sqrt\alpha$.

The total Hamiltonian is
\begin{equation} \label{ham}
H = T+\alpha V_c +H_f \ ,
\end{equation}
where the three terms are the kinetic energy of the particles, the
Coulomb potential energy and the quantized field energy, which will be
explained in detail presently.

The partition function is given by the trace $\Z=\tr \exp[-\beta H]$
and the pure-field partition function is given by $\Z_0= \tr
\exp[-\beta H_f]$, with $\beta = 1/k_BT$. We are interested in the
free energy per unit volume
\begin{equation}\label{freeen}
f= -k_B T \lim_{|\Omega| \to \infty,}\, \lim_{L\to \infty} \frac{1}{|\Omega|}
\left\{\log \Z -\log \Z_0 \right\}\ ,
\end{equation}
with the understanding that we set $N= \rho_{\mathrm{electron}} 
\cdot |\Omega|$ 
and $K=  \rho_{\mathrm{nucleus}}\cdot |\Omega|$ for some
fixed densities $\rho_{\mathrm{electron}}$ and $\rho_{\mathrm{nucleus}}$. 
We denote them, collectively, simply as $\rho$. Charge neutrality is not
assumed.

Our goal here is to derive a lower bound to $f$. We do not claim to
prove that the limits in (\ref{freeen}) exist. For the present
purpose they are interpreted as $\limsup$
instead of $\lim$.

We now define the various energies in detail. First, the kinetic
energies (in units of $2mc^2$).
\begin{equation}\label{kin}
T=  T^{\rm el}+T^{\rm nuc} = \sum_{i=1}^N T^P_i(A) +
\frac{1}{M}\sum_{j=1}^K T_j(-ZA)\ .
\end{equation}
The electron kinetic energy operator for each electron is 
the Pauli operator 
\begin{equation}\label{pauli}
T^P(A) = \left(\sigma\cdot(p+\sqrt{\alpha}\,  A(x))\right)^2 = 
(p+\sqrt{\alpha}\, A(x))^2
 + \sqrt{\alpha} \, \sigma \cdot B(x) \ ,
\end{equation}
which is appropriate for a spin 1/2 fermion in the presence of a
magnetic vector potential $A(x)$ and magnetic field $B(x)=
\mathrm{curl} A(x)$.  The operator $p$ is given (in our units) by
$p=-i{\mathbf{\nabla}}$.  The subscript $i$ in $T^P_i(A)$ in
(\ref{kin}) indicates that this operator acts on the coordinates of
electron $i$ and the $x$ in (\ref{kin}) is then $x_i$. Note that in
this model the $g$-factor of the electron is 2.  If it were greater
than this we would be in serious trouble because then the stability of
matter would not hold \cite{FLL}. (Strictly speaking the result in
\cite{FLL} about $|g|>2$ holds only for classical fields without UV
cutoff.  With a cutoff one expects stability of the first kind, i.e.,
a finite ground state energy, but not stability of the second kind,
i.e., a lower bound that is proportional to the number of particles.
Although well known QED calculations say that the renormalized,
effective $g$-factor exceeds 2, QED theory always starts with 2,
otherwise the theory would not be renormalizable \cite{Wei}.)

Since the nuclear charge is $+Z\sqrt{\alpha}$ we have $-ZA$ in
(\ref{kin}). The kinetic energy operator in (\ref{kin}) omits the
$\sigma \cdot B(x)$ term, i.e., it is
\begin{equation}
T_j(A)= (p_j+\sqrt{\alpha}\, A(R_j))^2  \  ,
\end{equation}
in which $R_j$ is the coordinate of the ${\mathrm{j}^{\mathrm{th}}}$
nucleus.  A nucleus can have a magnetic moment, even if its charge is
zero (the neutron) but it often has a $g$-factor much larger than 2
(e.g, $g\approx 5.5$ for a proton).

A conventional `physical argument' might be 
that since the magnetic moment is inversely 
proportional to the mass, the contribution of the magnetic energy
to the total energy is small.  As mentioned before, however,  the inclusion
of a dipole-dipole interaction has disastrous consequences, 
no matter how small the coupling is. It follows that to include this magnetic
interaction, and hence to include the hyperfine interaction, we would
have to take explicit account of the nuclear magnetic form factor. In
essence this really means thinking of the nuclei as a compound system
of quarks. The interaction of quarks with the EM field will surely
have diamagnetic components.  The effects of the nuclear structure,
which would add a self-energy proportional to $K$, could be included,
but we prefer to avoid this discussion here.

We use the Coulomb gauge to describe the EM field and its energy. In
this gauge only the magnetic field is a dynamical variable, i.e., the
curl-free part of the electric field is not an independent dynamical
variable, for it is determined by the particle coordinates and
Coulomb's law.  This choice of gauge is essential because, as we have
said elsewhere \cite{L, LLoss}, it is the only gauge in which the
correct physical EM interactions (including the spin-spin magnetic
interaction) can be obtained by a minimization principle.

The field energy (in our units) is 
\begin{equation}\label{fielden}
H_f =  \sum_k \sum_{j= 1}^3 |k| \, a^*_j(k)a_j (k) \ ,
\end{equation}
where the three operators $\mathbf{a}(k)= \left(a_1(k),\, a_2(k),\,
  a_3(k)\right)$ are boson annihilation operators of momentum $k$
satisfying the canonical commutation relations $\left[ a_{i} (q)\, ,\,
  a^*_j(k) \right]= \delta_{i,j} \delta_{q,k}$, etc. (We are using the
convention of three-component quantized fields introduced in
\cite{LLoss}, which means that we have to subtract the Planck
background energy in (\ref{freeen}) for {\it three} modes instead of
two. The advantage of this formalism is that we do not have to
introduce the $k$-dependent classical polarization vectors
$\varepsilon_\lambda(k)$ to insure the `divergence-free condition' on
the vector potential $A(x)$.) The $k$ sum in (\ref{fielden}) is over
$k$'s of the form $k=(2\pi/L)(n_1,\, n_2,\, n_3 )$ with integer $n_i$,
but $k=0$ is excluded.  As $L\to \infty $,
$(2\pi/L)^3$ times the sum
over $k$ becomes an integral $\int_{\R^3} $, but for a finite $L$ we
have to do the sum carefully in order to get the correct cancellation.

The vector potential is obtained by first defining the vector field
\begin{equation}\label{see}
C(x) =\frac{1}{2\pi}\left(\frac{2\pi}{L}\right)^{3/2}  \sum_k \chi^{\phantom 
*}_\Lambda(k) 
\frac{1}{|k|^{3/2}}\left[
\mathbf{a}(k) e^{ik\cdot x} - \mathbf{a}^*(k) e^{-ik\cdot x}\right]\ ,
\end{equation}
where $\widehat{\chi}_\Lambda \leq 1$ is a radial function that 
vanishes outside a 
ball of radius $\Lambda$.
Then 
$$
A(x) = i\, \mathrm{curl} \, C(x)\ .
$$

In the Coulomb gauge the electrostatic  energy $\alpha V_c$ 
is given by a simple
coordinate-dependent potential
\begin{equation}\label{coulomb}
V_c = \sum_{1\leq i < j\leq N} \frac{1}{|x_i-x_j|} 
+Z^2 \sum_{1\leq i < j\leq K} \frac{1}{|R_i-R_j|} 
-Z \sum_{1\leq i \leq N}  \sum_{1\leq j \leq K} \frac{1}{|x_i-R_j|}\ .
\end{equation}

Having introduced the free energy and its component parts we can now
discuss the physical and mathematical problems addressed in this paper,
namely the difference between the QED problem and the non-QED problem in
\cite{LLeb}. 

\medskip
\quad {\bf The non-QED problem:} In this case there is no need to introduce the
`universe' $\mathcal{U}$, the field energy $H_f$ or to take the double limit in
(\ref{freeen}) because there is no particle-field interaction via the
field $A(x)$. To obtain a lower bound to $f$ one simply writes
\cite[Theorem 2.2]{LLeb} $H=\widetilde{H}_1 +\widetilde{H}_2$ with
\begin{equation}\label{h12}
\widetilde{H}_1 = \frac{1}{2}T + \alpha V_c  \quad\quad\quad\quad 
\widetilde{H}_2 = \frac{1}{2}T\  .
\end{equation}
One then bounds $\widetilde{H}_1$ from below by $-c (N+K)$, where $c$ is a universal 
constant,
which follows from the
stability of matter bound. (The constant $c$ changes when we replace 
$T$ by $T/2$ but it
is always finite.)
Then, we can bound $f$ by
\begin{equation}\label{fdecompose}
f \geq  -c\left(\rho_{\rm electron} +\rho_{\rm nucleus}\right)
 -k_BT \lim_{\Omega\to \infty} 
|\Omega|^{-1}\log \tr \,e^{-\beta T/2}\ .
\end{equation}
The latter trace is just the partition function of an ideal gas of twice the
mass and has, as the elementary textbooks tell us, a finite $f$ (which
can be easily bounded from below by a shape independent $f$).  Thus, a
satisfactory lower bound to $f$ is a `two-line argument' in this case.

\medskip

\quad {\bf The QED problem:} Several points have to be considered.
\begin{enumerate}
\item The precise cancellation of the background Planck field energy, 
which is enormously greater than $f$, has to be done
carefully. \label{item1}

\item We know from earlier work \cite{LLS} that the stability of
  matter does not hold for the Pauli kinetic energy unless the field
  energy is added to the Hamiltonian.  This implies that we must
  somehow borrow some field energy to stabilize the Pauli analog of
  $H_1 $ above, but not too much to spoil the delicate cancellation in
  item \ref{item1}.  The situation with the quantized field energy
  $H_f$ in place of a classical field energy $\int_{\R^3}B(x)^2 dx
  /8\pi$ is even more delicate; this extension was first made in
  \cite{BFG} using the results in \cite{LLS}. The idea in \cite{BFG}
  is to replace the field energy term by one that is localized near
  the nuclei.  Our approach will be somewhat different and will only
  involve localization of the electrons, in a manner reminiscent of
  the original Dyson-Lenard proof \cite{DL} of stability of matter.
  (However, if we do not care that our lower bound for $f$ does not go
  to zero as $\rho \to 0$ then localization is not needed -- see
  section \ref{simpler}.)  In any event, the stability of matter using
  the Pauli operator and $H_f$ {\it requires a bound on $\alpha$ and
    $Z\alpha^2$,} when there is no ultraviolet cutoff; this is a
  feature not encountered in the non-QED problem. We shall prove that
  $f$ is bounded below (for all $\Lambda$ if $\alpha \leq 1/98$ and if
  $Z\alpha^2 <1/468$ (corresponding to $Z\leq 39$ for $\alpha =
  1/137$.  These bounds can certainly be improved, with some effort,
  but we shall not attempt to do so.

\item The lower bound on $f$ should be shown to have a classical limit
  as the nuclear mass $M$ tends to infinity, independent of the
  statistics of the nuclei. `Classical' means that the dependence of
  $f$ on $M$ has the form $f \sim K \rho_{\rm nucleus} k_BT
  \left[\frac{3}{2}\log(\beta/M) + \log \rho_{\rm nucleus} -1 \right]$.
    This would coincide with the experimental observation that the
    nuclei may as well be considered to be fixed in space.  \label{class}
    
\item While an infrared cutoff is {\it not} needed for our considerations,
    the ultraviolet cutoff $\Lambda$ is essential. One would like to
    take the limit $\Lambda \to \infty$ after a renormalization of the
    electron mass and possibly its charge. At present, it is not known
    how to carry out this program, although some primitive steps were
    taken in \cite{LLrenorm}. In any case, in order to show that
    $\Lambda \to \infty$ limit can be taken for the thermodynamics it
    is appropriate and necessary to show the following.  There are three
    functions $f_\infty (\beta, \rho, Z) $, 
    $g_{\rm nucleus} (\Lambda, Z)$ and $g_{\rm electron}(\Lambda)$
    such that the free energy per unit volume, $f(\beta, \rho, Z, \Lambda)$,
    which depends on all four parameters, can be decomposed as
\begin{equation}\label{freendecomp}
f(\beta, \rho, Z, \Lambda) = f_\infty (\beta, \rho, Z) + g_{\rm
electron}(\Lambda)\,
\rho_{\rm electron} +g_{\rm nucleus}(\Lambda, Z)\, \rho_{\rm nucleus}  \ .
\end{equation}
We make the additional requirement that $f_\infty (\beta, \rho, Z)
\propto \rho$ as $\rho \to 0$. With no particles there should be no 
free energy. (It is also physically desirable that the self-energy term
$g_{\rm electron}(\Lambda)$ should be large and positive when
$\Lambda$ is large. We succeed only
partially in this respect, as discussed in the remark following
Theorem \ref{mainthm}.)
\end{enumerate}

In this paper a lower bound of the form (\ref{freendecomp})
will be derived and it will have the indicated properties.
This is done is in section \ref{strategy} with help from
section \ref{boxes} and the Appendices

If we forego the property that  $f_\infty (\beta, \rho, Z)
\propto \rho$ as $\rho \to 0$ then it is
significantly easier to obtain a lower bound of the form
(\ref{freendecomp}). This is shown in section \ref{simpler}.

%%%%%%%%%%%%%%%%%%%%%%%%%%%%%%%%%%%%%%%%%%%%%%%%%%%%%%%%%%%%%%%%%%%%
%%%%%%%%%%%%%%%%%%%%%%%%%%%%%%%%%%%%%%%%%%%%%%%%%%%%%%%%%%%%%5

\section{Main Theorem and Proof} \label{strategy}

In this section we show how to decompose $H$ in a manner reminiscent
of (\ref{h12}) and how to use this to prove a lower bound of the form
(\ref{freendecomp}), under the assumption that certain inequalities
hold.  These inequalities will be proved in subsequent sections and
appendices.  In the following $\Z_{\rm nucleus}$ is the partition
function of non-interacting bosons of mass $Mm$.  As $M \to \infty$
this partition function tends to the classical partition function if
$T>0$. This was one of our requirements in item \ref{class} in section
\ref{sec2}.

Thus, we shall prove the following in this section.

\begin{thm}[Lower bound on f] Assume that 
$\alpha < 1/98$ and $Z\alpha^2 < (\frac{39}{40})(\frac{1}{468})$. 
(This means 
$Z\leq 39$ when $\alpha = 1/137$.) Then
(\ref{fdecompose}) holds with
\begin{align} 
f_\infty(\beta, \rho, Z) &> -kT\left\{\rho_{\rm electron}\ln 2 
+\rho_{\rm electron}\, \ln (\rho_{\rm electron})
 -\rho_{\rm electron} \right\}  \nonumber \\
&\quad\quad - kT 
\rho_{\rm electron} \, \ln \left( 8\pi (kT)^3C_3^{-3}\right)
 +f_{\rm nucleus}(\beta,  \rho_{\rm nucleus})
\  ,   \label{mainbound1}\\
& \nonumber \\
g_{\rm electron}(\Lambda) &> 
 -\left( 128  +0.0032 \right)\Lambda^{8/5}
   - \left(\frac{1}{149\alpha}\right)^2  
   -(\frac{4\pi}{3})(468)(40) \Lambda^{3/5}, \\
g_{\rm nucleus} (\Lambda, Z) &\geq  0
\ , \label{mainbound3}
\end{align}
where $f_{\rm nucleus}$ is the free energy per unit volume for
non-intercting 
bosonic nuclei of mass $mM$ and
\begin{equation}
C_3\alpha = \frac{1}{149} -\alpha^2 \max \{64.5,\, \pi Z\}
-\frac{1}{149}\exp\left\{\frac{-468}{1-468Z\alpha^2}\right\}
\end{equation}
 \label{mainthm}
\end{thm}
See equation (\ref{bound}) for a more general expression.

{\it REMARK:\/} The bounds (\ref{mainbound3}) on $g_{\rm
electron}(\Lambda)$ and $g_{\rm nucleus} (\Lambda, Z)$ should actually be
large and positive owing to the self-energy of the charged particles. In
\cite{selfenergy} we showed that the dependence of this  self-energy
on $\Lambda$ is somewhere between $\Lambda^{3/2}$ and $\Lambda^{12/7}$
and we conjectured that $\Lambda^{12/7}$  was the correct dependence.
If  $\Lambda^{12/7}$  is, indeed, correct then we could easily invoke
the methods of    \cite{selfenergy} and add a term proportional to
$\Lambda^{12/7}$ to  $g_{\rm electron}(\Lambda) $, which would dominate
the $\Lambda^{8/5}$ term and thereby leave us with the desired large
positive value for $g_{\rm electron}(\Lambda) $.  On the other hand,
if $\Lambda^{3/2}$ is correct then adding $\Lambda^{3/2}$ to $g_{\rm
electron}(\Lambda) $ will not produce a positive self-energy (for large
$\Lambda$, of course). Since the $\Lambda^{3/2}$ --  $\Lambda^{12/7}$
question is not yet settled, we shall not burden this paper further
with the  question of the sign of the self-energy.  We note in passing,
however, that the simpler theorem in section \ref{simpler} has a huge
negative $\Lambda$ dependence (i.e., $-\Lambda^4$), which cannot be
compensated even by $\Lambda^{12/7}$.

\begin{proof}
The first step is to localize the electrons. Decompose $\Omega$ into
disjoint cubes of sidelength $\Lambda^{-s}$. There will be
approximately $|\Omega|\, \Lambda^{3s}$ of these cubes.  For each cube
$\gamma_\ell$ we take a smooth function $\phi_\ell$ centered on
$\gamma_\ell$, whose support is in a cube $\Gamma_l$ of twice the side
length and in such a way that
\begin{equation}\label{norm}
\sum_\ell \phi_\ell (x)^2 =1 \quad\quad {\mathrm{for\ all}}\ x\in \Omega \ .
\end{equation}
We can and do require  that $|\nabla \phi_\ell(x)| \leq 4\Lambda^s$.

It is easily seen that for each point $x$ there can be at most $8$
distinct values of $\ell $ for which $ \phi_\ell (x) \neq 0$, and
hence, by the standard IMS localization formula,
\begin{equation}
T^P(A) =  \sum_\ell \phi_\ell(x) T^P(A)  \phi_\ell(x) -  \sum_\ell 
|\nabla\phi_\ell(x)|^2
\geq \sum_\ell \phi_\ell(x) T^P(A)  \phi_\ell(x) - 128\Lambda^{2s} \ .
\end{equation}
for all $A(x)$.
By applying this to each term in $ T^{\mathrm{electron}} $, 
and recalling (\ref{norm}),
we obtain the 
inequality
\begin{equation} \label{kinloc}
 T^{\mathrm{electron}} \geq \sum_I \Phi_I  T^{\mathrm{electron}} \Phi_I -
128 N\Lambda^{2s }\ ,
\end{equation}
where $I = (\ell_1, ... , \ell_N)$ is a multi-index and
$\Phi_I({\underline{X}}) = \Phi_I (x_1, ... , x_N)
=\phi_{\ell_{1}}(x_1) \cdots \phi_{\ell_{N}}(x_N)$ is a product
function, which satisfies $\sum_I \Phi_I({\underline{X}})^2 =1$ when
${\underline{X}} \in \Omega^N$. Here, ${\underline{X}}= (x_1, x_2,
\cdots , x_N)$ collectively denotes the $N$ electron coordinates;
similarly, ${\underline{R}}= (R_1, R_2, \cdots , R_K)$ denotes the $K$
coordinates of the nuclei.

Armed with this localization, and recalling (\ref{norm}), we can write
(for an arbitrary constant $C_1>0$)
\begin{multline}\label{tbound}
 T^{\mathrm{electron}} \geq \sum_I \Phi_I({\underline{X}}) \
\sum_{i=1}^N \left\{T^P_i(A) -
C_1|p_i+\sqrt\alpha A(x_i)| \right\} \Phi_I ({\underline{X}}) \\ + \sum_I 
\Phi_I({\underline{X}}) 
\sum_{i=1}^N 
C_1|p_i+A(x_i)| \Phi_I({\underline{X}}) - 128 N \Lambda^{2s}\ ,
\end{multline}
and hence our Hamiltonian is bounded below as
\begin{equation}
H\geq H_2 :=\sum_I \Phi_I({\underline{X}}) \left[H^{\mathrm {rad}} + H^{\mathrm 
{rel}}\right]\Phi_I ({\underline{X}}) 
-128 N \Lambda^{2s} \ ,
\end{equation}
where
\begin{align}
H^{\mathrm {rad}} &= \sum_{i=1}^N \left\{T^P_i(A) - C_1|p_i+\sqrt\alpha 
A(x_i)| \right\} +H_f \label{Hrad} \\
H^{\mathrm {rel}} &=  T^{\mathrm{nucleus}} + C_1\sum_{i=1}^N |p_i+\sqrt\alpha 
A(x_i)| + \alpha V_c \ . \label{rel}
\end{align}
The superscript `rel' is meant to suggest that (\ref{rel}) is a
Hamiltonian of relativistic-like electrons and non-relativistic
nuclei. Note that $\sum_I\Phi_I(\underline{X}) H_f
\Phi_I(\underline{X}) = H_f$ since $H_f$ does not depend on the
electron coordinates ${\underline{X}}$.

The reason for adding and subtracting the `relativistic' kinetic
energy operator $|p+\sqrt{\alpha}A(x)|$ is that $\tr \exp\{-T^P(A)\}$ is
not bounded independent of the vector field $A$, but
$\tr \exp\{-|p+\sqrt{\alpha}A(x)|\}$ is uniformly bounded.

In sect. \ref{boxes}, Lemma \ref{boxenergy}, we shall show that the sum 
appearing in 
(\ref{Hrad}) is bounded below by
\begin{equation}\label{tbound2}
 - \frac{16 \alpha C_1\mathcal{L} }{3\sqrt 3}\sum_I
\Phi_I({\underline{X}})^2 \int_{\R^3} B(y)^2 W_I(y) dy
-N C_1^2\ ,
\end{equation}
where $W_I$, given in (\ref{w}), is a sum of characteristic functions
of subsets of $\Omega$ whose total volume is at 
most $8N\Lambda^{-3s}$. This function has the following properties for
each $I$:
\begin{equation}\label{wprop}
\int_{\R^3}W_I(y)dy \leq 8N\Lambda^{-3s}, \quad\quad W_I(y)\leq 8 \ .
\end{equation}
The constant $\mathcal{L}$ stems from the application of a 
Lieb-Thirring inequality and it is bounded above by  $0.06003$.

If $D(y)$ is the (vector) operator obtained from
the $\mathbf{a}$ part of (\ref{see}), namely
\begin{align}
D(y) &= \frac{i}{2 \pi}\, \mathrm{curl}\ \mathrm{curl} \, 
\left(\frac{2\pi}{L}\right)^{3/2}\sum_k \chi^{\phantom *}_\Lambda(k) 
\frac{1}{|k|^{3/2}}
\mathbf{a}(k) \, e^{ik\cdot y} \\
&=\frac{1}{2\pi}\left(\frac{2\pi}{L}\right)^{3/2}\sum_k  
\chi^{\phantom *}_\Lambda(k) \frac{1}{|k|^{3/2}}
 e^{ik\cdot y}\left[k^2 \mathbf{a}(k) 
- (k\cdot   \mathbf{a}(k)) k \right]           \ ,
\end{align} 
then $B(y)=D(y)+D^*(y)$ and Schwarz's inequality leads to
\begin{equation}
B(y)^2 \leq 4 D^*(y)D(y) + 2 \left[D(y), D(y)^*\right] \ ,
\end{equation}
which then implies that
\begin{equation}\label{commutator}
\int_{\R^3} B(y)^2 W_I(y) dy \leq 4\int_\Omega D^*(y)\cdot  D(y) 
W_I(y) dy +
 \frac{8}{\pi} \Lambda^{4-3s}N \ .
\end{equation}

Altogether, (\ref{tbound2}) and the definition of $H^{\mathrm {rad}}$ and 
$H^{\mathrm {rel}}$
lead to the lower bound
\begin{equation}\label{H3}
H\geq H_2 \geq H_3 := \sum_I \Phi_I({\underline{X}}) \left[H_1^{\mathrm {rad}} + 
H^{\mathrm {rel}}\right]\Phi_I ({\underline{X}}) 
-\left[128  \Lambda^{2s} + C_1^2+ \frac{128 \alpha C_1 
\mathcal{L}}{3 \sqrt 3 \pi} \Lambda^{4-3s}\right] N \ ,
\end{equation}
where $H_1^{\mathrm {rad}}$ is a replacement for $H^{\rm rad}$ given by
\begin{equation} \label{rad}
H_1^{\mathrm {rad}}= H_f - \frac{64 \alpha C_1 \mathcal{L}}{3\sqrt 3} \sum_I
\Phi_I({\underline{X}})^2 \int_{\R^3} D^*(y)D(y) W_I(y) dy \ .
\end{equation}

{\it In summary, our lower bound Hamiltonian $H_3$ contains three
    parts: A constant proportional to $N$, a perturbed field energy
    $\sum_I \Phi_I H_1^{\rm rad}\Phi_I$, and the Hamiltonian $\sum_I
    \Phi_I H^{\rm rel}\Phi_I$ of `relativistic' electrons and
    non-relativistic nuclei.  Their definitions depend on a constant
    $C_1>0$ and on the parameter $s$, which defines the electron
    localization. These will be chosen later.}

Our final goal is to prove the following upper bound on $\tr \,
\exp\{-\beta H_3\}$, which will then complete the proof of
Theorem \ref{mainthm}.

\begin{lm} \label{gothomp}
\begin{multline}\label{lem32}
\tr \, e^{-\beta H_3}  \leq  \Z_{\rm nucleus} (K) \frac{1}{N!}
\left[ \int_{\R^3} e^{-\beta C_3|p|} d p \right]^N 
\ 
{\rm Tr}_{\mathcal F}\, e^{-\beta H_f} \times \\  
\exp\left\{\beta N (128 \Lambda^{2s} + C_1^2 + \frac{128 \alpha C_1 \mathcal{L} }
{3 \sqrt 3 \pi} \Lambda^{4-3s}+ \frac{4 \pi}{3}|\ln(1-\varepsilon)|
\Lambda^{3-3s})\right\} \ ,
\end{multline} 
where $\Z_{\rm nucleus} (K)$ is the partition function of $K$
non-interacting bosons of mass $M$ (even if some of the nuclei happen
to be fermions).  The number $\varepsilon$ (assumed to be $<1$, see
(\ref{eps2})) is
\begin{equation}\label{epsilon}
\varepsilon = \frac{4096\, \pi}{3\sqrt 3}  \alpha C_1 \mathcal{L}.
\end{equation}
\end{lm}

\begin{proof}
First, we dispose of the localization function $\Phi({\underline{X}})$
that appear in the Hamiltonian $H_3$ of (\ref{H3}).
By Lemma \ref{loc} in Appendix B, and the fact that 
$\sum_I \Phi_I(\underline{X})^2=I=$ identity operator
on our Hilbert space (\ref{hilbert}),

\begin{equation}
{\rm Tr}\,  e^{-\beta H_3 } \leq {\rm Tr}\, 
e^{-\beta (H_1^{\mathrm {rad}} +H^{\mathrm {rel}})}  \, 
\exp \left\{\beta N (128 \Lambda^{2s} + C_1^2 + 
\frac{128 \alpha C_1 \mathcal{L} }{3 \sqrt 3 \pi} \Lambda^{4-3s} )\right\} \ .
\end{equation}

Next, we introduce another constant $0< C_3 < C_1$, and write
\begin{equation}
H^{\mathrm {rel}} =  T^{\mathrm{nucleus}} + C_3 \sum_{i=1}^N |p_i+\sqrt\alpha 
A(x_i)|
+\sum_{i=1}^N (C_1-C_3)|p_i+\sqrt\alpha A(x_i)| + \alpha V_c \ .
\end{equation}
Using the result of \cite{LY}, as stated in \cite{LLS} (see also 
\cite{LLSie}), the last two terms, taken together, are  positive
as an operator on the tensor product space of the nuclei and the 
spin $1/2$ electrons 
provided that
\begin{equation}\label{condone}
{C_1-C_3} \geq \max\{ \frac{2}{0.032}, \pi Z\} \alpha \ .
\end{equation}
If (\ref{condone}) is true then
\begin{equation}
H^{\mathrm {rel}} \geq H_1^{\mathrm {rel}}: = T^{\mathrm{nucleus}} + 
C_3 \sum_{i=1}^N 
|p_i+\sqrt\alpha A(x_i)| \ .
\end{equation}

Using this and the Golden-Thompson inequality
\begin{equation} \label{gt}
{\rm Tr}\, e^{-\beta (H_1^{\mathrm {rad}} +H^{\mathrm {rel}})}
\leq {\rm Tr}\, e^{-\beta (H_1^{\mathrm {rad}} +H_1^{\mathrm {rel}})}  
\leq {\rm Tr}\left(e^{-\beta H_1^{\mathrm {rad}}}
e^{-\beta H_1^{\mathrm {rel}}}\right) \ .
\end{equation}

We evaluate the trace in the Schr\"odinger representation in which
the field $A$ is a c-number field and the trace is just integration
over this classical field. This is a rigorous technique in quantum
field theory and we explain it in some detail in Appendix \ref{schroe}.
We then use the
fact that for a fixed classical field $A(y)$ the operator $e^{-\beta  
H_1^{\mathrm {rel}}}$
has a kernel $\exp \{-\beta H_1^{\mathrm {rel}}\}(\underline X, \underline R\ ;\ 
\underline X, \underline R)$, and we can write the right side of
(\ref{gt}) as
\begin{multline} \label{separated}
\tr e^{-\beta  H_1^{\mathrm {rel}}} e^{-\beta H_1^{\mathrm {rad}}}=
2^N\, \int \mathcal{D} 
(A) \int d\underline X 
d \underline R \\ e^{-\beta H_1^{\mathrm 
{rel}} (A)}(\underline X, 
\underline R, \underline  \ ;\ 
\underline X, \underline R, \underline ) \ 
\langle A|\, e^{-\beta H_1^{\mathrm {rad}}}\, 
| A\rangle 
(\underline X ) \ ,
\end{multline}
where the factor $2^N$ comes from the electron spins. (Note that $H_3$
has no electron-spin dependence.) There might be a factor for nuclear
spins, but we shall ignore this subtlety in order to keep the notation
simple.
In any case it can be absorbed in the  factor $\Z_{\rm nucleus}(K)$.

Obviously, the matrix element $e^{-\beta H_1^{\mathrm
{rel}}(A)}(\underline X, \underline R\ ;\ \underline X, \underline
R)$ is the product of a factor depending on the nuclear coordinates
$\underline R$ and a factor depending on the electron coordinates
$\underline X$. Each of these factors is an ($A$-dependent)
partition function of non-interacting particles. The former
depends on the statistics of the nuclei. If the nuclei are bosons
$e^{-\beta H_1^{\mathrm {rel}}(A)}(\underline X, \underline R\ ;\ 
\underline X, \underline R)$ can be written as
\begin{equation}
\frac{1}{K!} {\rm per} \left(\exp\{-\frac{\beta}{M}(p- Z\sqrt \alpha \,
A(x))^2\}(R_j, R_k) \right) \times
\frac{1}{N!}  \det \left(\exp\{-\beta C_3|p+\sqrt \alpha A(x)|\, \}
(x_i, x_j)\right) \ ,
\end{equation}
where ${\rm per}$ indicates the permanent of the $K\times K$ matrix
(indexed by $j,k\in \{1,\dots ,K\}$) 
and det the determinant of the $N\times N$ matrix 
(indexed by $i,j\in \{1, \dots ,N \}$).  If the
nuclei are also fermions or a mixture of fermions and bosons we have
to replace the permanent above by the corresponding symmetrized or
antisymmetrized product, i.e., permanent for a boson species or
determinant for a fermion species.

For our purpose here, namely an upper bound, we may assume from now on
that all the nuclei are bosons. The reason is that the $K\times K$
matrix above is positive definite, and so is the $N\times N$ matrix.
It is a fact that the determinant of a positive definite matrix is not
greater than the permanent. Indeed, the determinant is less than or
equal to the product of the diagonal entries while the permanent is
greater than or equal to 
the same product.  (See (\cite{perm}).)
Since $\langle A | e^{-\beta H_1^{\rm rad}}| A \rangle $ is positive,
we can use this upper bound on the determinant to obtain the following
upper bound to the right side of (\ref{separated})
\begin{multline}\label{separ}
\int{\mathcal{D}}(A)  \int d\underline X \, d \underline R \frac{1}{K!}
 {\rm per} 
\left(e^{-\frac{\beta}{M}(p- Z\sqrt \alpha A(x))^2}(R_j, R_k) 
\right)\\            
\times \frac{1}{N!}  \prod_{i=1}^N \left(e^{-\beta C_3|p+\sqrt \alpha A(x)|}
(x_i, x_i)\right) 
\times \langle A \, | e^{-\beta H_1^{\mathrm {rad}}}\,  |A\rangle 
(\underline X, ) \ ,
\end{multline}

Since all the factors in (\ref{separ}) are positive, we can appeal to
the diamagnetic inequality and delete the field $A$ from the second
factor. (The diamagnetic inequality is well known and states that
$\exp\left[-\beta |p+\sqrt \alpha A(x)| ^2\right](x_i,
x_j)\leq\exp\left[-\beta |p|^2\right](x_i, x_j) $, which follows from
the Feynman-Kac representation; it is also true for $\exp \left[-\beta
  |p+\sqrt \alpha A(x)| \, \right](x_i, x_j)$, thanks to the fact that
$ e^{-|p|} = \int_0^\infty e^{-t-p^2/4t}dt/\sqrt{\pi t}$.) Note that
it was first necessary to replace the determinant by the product of
its diagonal elements and then to use the diamagnetic inequality;
otherwise we would have to worry about the minus signs in the
determinant

Similarly, we can set $A=0$  in the first (permanent) factor. The
reason is  that the  permanent can  only increase if  we replace  each
matrix element by its  absolute value and then replace that, in turn, by
a larger number. But the diamagnetic  inequality  (actually, the Wiener
integral representation, to be precise) tells us
that this is achieved by setting $A=0$ (even if $R_j\neq R_k$).

In this manner we obtain the upper bound
\begin{equation}
 \int d\underline X \, d \underline R \frac{1}{K!} {\rm perm} 
\left(e^{-\frac{\beta}{M}p^2}(R_j, R_k) \right)\times
\frac{1}{N!}  \prod_{i=1}^N \left(e^{-\beta C_3|p|}(x_i,x_i)\right)  
{\rm Tr}_{\mathcal F}e^{-\beta H_1^{\mathrm {rad}}} (\underline X, ) 
\ .
\end{equation}
The expression containing the Fock space trace still depends
on the variables $\underline X$. Using
(\ref{energydifference}) from Appendix \ref{radiation} we see that
\begin{equation}\label{eps}
\ln {\rm Tr}_{\mathcal F}e^{-\beta H_1^{\mathrm {rad}}} (\underline X, 
\underline R) \le 
\ln {\rm Tr}_{\mathcal F}e^{-\beta H_f} + \frac{4\pi  }{ 3} \left| \ln
(1-\varepsilon)\right| \Lambda^{3-3s} \ ,
\end{equation}
where $\varepsilon$ is given in (\ref{epsilon}) and where (see (\ref{wprop}))
\begin{equation}\label{G}
G(y, \underline{X}) = \sum_I  \Phi_I(\underline {X})^2 W_I(y) \leq 8 \ .
\end{equation}
Inequality (\ref{eps}) is true, as shown in (\ref{positivity}),
(\ref{energydifference}) provided the criterion $M \leq \varepsilon K$
given there is satisfied.  Since $ G(y, \underline{X}) \leq 8$ for all
$y, \underline{ X}$ this criterion is satisfied with 
$\varepsilon $ as in (\ref{epsilon}),
and this can be achieved by choosing $C_1$ small enough.

Recalling that the operators $p^2$ (associated with the nuclei) and
$|p| = \sqrt{p^2}$ (associated with the electrons) are Dirichlet
Laplacians on the domain $\Omega$ we get the upper bound
\begin{multline}
\tr \, e^{-\beta H_3}  \leq  \Z_{\rm nucleus} (K) \frac{1}{N!}
\left [{\rm Tr}\, e^{-\beta \sqrt{p^2}}\right]^N \ 
{\rm Tr}_{\mathcal F}\, e^{-\beta H_f}\\  
\exp\left\{\beta N (128 
\Lambda^{2s} + C_1^2 + \frac{256 \alpha \mathcal{L} }{3 \sqrt 3 \pi}
C_1 \Lambda^{4-3s}+\frac{4 \pi}{3}|\ln (1-\varepsilon)|  
\Lambda^{3-3s})\right \} \ .
\end{multline}
This proves Lemma \ref{gothomp} 
\end{proof}

 The  factor 
$\frac{1}{N!} \left[{\rm Tr}\, e^{-\beta C_3 \sqrt{p^2}}\right]^N $
can be estimated from above, by the Golden-Thompson inequality, as
\begin{equation}\label{kinpart2}
\frac{1}{N!}\left[ \int_{\R^3} e^{-\beta C_3|p|} d p \right]^N |\Omega|^N \ ,
\approx
\exp\left\{ |\Omega | \left( -\rho_{\rm electron}\, \ln (\rho_{\rm electron})
 +\rho_{\rm electron} +
\rho_{\rm electron} \, \ln (8\pi \beta^{-3} C_3^{-3}) \, \right) \right\}
\end{equation}

To prove Theorem
\ref{mainthm} we have to consider numerical values for our constants.
Let us collect together the conditions on them, which are
(\ref{epsilon}), (\ref{condone}). That is
\begin{align}\label{conditions}
\varepsilon &= \frac{(4096)(0.06)\, \pi}{3\sqrt{3}}\alpha C_1 
= 149 \, C_1\alpha <1, \\
C_3 &= C_1 - {\mathrm{max}}\{\frac{2}{0.032},\, \pi Z\}\alpha 
= C_1 - {\mathrm{max}}\{64.5, \, \pi Z\}\alpha>0 \ .
\label{see3}
\end{align}
This value of $C_3$ is to be inserted into \ref{lem32}, using
(\ref{kinpart2}) --- assuming that the two conditions on $C_1 $,
implied by
(\ref{conditions}) and (\ref{see3}), are satisfied. These two conditions set bounds
on $\alpha$ and on $Z\alpha^2$. These are
$(149)(64.5)\alpha^2 =
9613\, \alpha^2 <1$ and $(149)\pi Z \alpha^2 = (468)\, Z\alpha^2 <1$,
as stated in Theorem \ref{mainthm}.   

The free constants to be determined are $C_1$ and  $s$. The other constants
$\varepsilon$ and $C_3$ are in (\ref{epsilon}) and (\ref{see3}), respectively.
The factor 
$128 \alpha C_1 \mathcal{L} 
/3 \sqrt 3 \pi$ in (\ref{lem32}) can be replaced by $0.47/149= 0.0032$ since
$C_1\alpha < 1/149$. Our bound is then
\begin{align}
-\beta f \leq &\rho_{\rm electron} \ln\, 2 -\beta f_{\rm nucleus}\\
&+\beta \rho_{\rm electron}\left(128 \Lambda^{2s} +C_1^2 +
0.0032  \Lambda^{4-3s} +\frac{4\pi}{3} |\ln(1-\varepsilon)|
\Lambda^{3-3s} \right) \label{selfen} \\
&-\rho_{\rm electron}\, \ln (\rho_{\rm electron})
 +\rho_{\rm electron} +
\rho_{\rm electron} \, \ln (8\pi \beta^{-3} C_3^{-3})\ ,
\label{bound}
\end{align}
where  $f_{\rm nucleus}=-kT|\Omega|^{-1} \ln\, \Z_{\rm nucleus}$ is the 
free energy per unit volume for non-interacting bosonic nuclei of mass
$mM$. 

If we choose, for example, 
\begin{equation}
\varepsilon= 149\, C_1\alpha =1 -\exp\left(\frac{-468}{1 -468\, Z\alpha^2}
\right),
\end{equation}
and $s=4/5$ in order that the two largest $\Lambda$ exponents in
(\ref{selfen}) have a common value (8/5), and  
restrict $Z\alpha^2 \leq (\frac{39}{40})(\frac{1}{468})$, 
then (\ref{mainbound1})-(\ref{mainbound3}) is obtained. 
\end{proof}

%%%%%%%%%%%%%%%%%%%%%%%%%%%%%%%%%%%%%%%%%%%%%%%%%%%%%%%%%%%%%%%%%%%%%

%%%%%%%%%%%%%%%%%%%%%%%%%%%%%%%%%%%%%%%%%%%%%%%%%%%%%%%%%%%%%%%%%%%%%%%%%%%%%%

\section{ Decomposition into Boxes} \label{boxes}

In this section we shall give the details of the lower bound,
(\ref{tbound2}), of the kinetic energy operator contained in
(\ref{tbound}) in terms of a Fock space energy operator.  We recall the
IMS localization into disjoint cubes $\gamma_\ell$ with side length 
$ \Lambda^{-s}$ and overlapping cubes $\Gamma_\ell$ with twice the side
length introduced in (\ref{norm}) -- (\ref{kinloc}).

\begin{lm}\label{boxenergy}
On the Hilbert space $\wedge_{i=1}^N L^2(\R^3;\C^2)$ 
of $N$ electrons with $2$ spin states we have that for
all values of $C_1 $ and all vector potentials $A(x)$, 
\begin{multline}\label{lowerbound}
 \sum_{I} \Phi_{I}(\underline X ) \sum_{j=1}^N 
\left\{T^P_j (A)-C_1|p_j+\sqrt \alpha A(x_j)| \right\}
\Phi_{I}(\underline X) \geq \\
-\frac{16 \alpha C_1 \mathcal{L}}{3 \sqrt 3} 
\sum_{I} \Phi_{I}(\underline X)^2 
\int_{\R^3} B(x)^2 W_I (x)dx  -C_1^2N\ ,
\end{multline}
where
the function $W_I(x)$, \ $x\in \Omega$, \  is given by
\begin{equation}\label{w}
W_I(x) = \sum_{k \in I} \chi_k(x) \leq 8 \ ,
\end{equation}
where $\chi_k$ is the characteristic function of the cube $\Gamma_k$
and where $k \in I =(\ell_1, \dots , \ell_N) $ means that at least one
of the $\ell_i $ equals $k$. Note our convention that each $\chi_k$ is
allowed to enter the sum in (\ref{w}) at most once, i.e., if $k$
appears 5 times in $I$ then $\chi_k$ appears once in (\ref{w}). The
constant $\mathcal{L}$ in (\ref{lowerbound}) is the $\gamma =1/2$,
3-dimensional Lieb-Thirring constant; $\mathcal{L} < 0.06003$.
\end{lm}

Note: See \cite{char} for the value of $\mathcal{L}$ quoted above and
see \cite[appendix A]{Loss} for the fact that it is not necessary to
include an extra factor of 2 in order to account for the 2 spin
states.
\begin{proof}
  Fix $I = (\ell_1, \dots, \ell_N)$ and consider the single term
\begin{equation}\label{y}
Y_I= \Phi_{I} (\underline X) \sum_{j=1}^N 
\left[ T_{j}^P(A)-C_1|p_j+\sqrt \alpha A(x_j)|\right]
\Phi_{I}(\underline X) \ . 
\end{equation}
In the index set $I$ the index $k_1$ appears $n_1$ times, the index
$k_2$ appears $n_2$ times etc. where the numbers $n_i \geq 1$ and
$\sum_i n_i= N$. 

Our goal is to find a lower bound to $(\Psi, Y_I \Psi) $ for any $\Psi
$ in $\wedge^N L^2(\Omega; \C^2)$.  Let us consider the first $n_1 $
terms in (\ref{y}), i.e., $ (\Phi_I\Psi, \sum_{j=1}^{n_1
}\mathcal{T}_j \ \Phi_I\Psi)$, where $\mathcal{T} $ is the operator
appearing in $[\ \ \ ]$ in (\ref{y}).  In evaluating this inner
product we can fix the coordinates $x_j, \sigma_j$ with $j= n_1 + 1,
\dots, N$ and then integrate over them at the end.  In other words,
the proof of our inequality (\ref{lowerbound}) will follow from the
following statement:
For each $n>1$ and each $k$, every normalized, antisymmetric function
$\psi$ of $n$ space-spin variables, with support in $\left(\Gamma_{k}\right)^n$
satisfies the inequality
\begin{equation} \label {thisisit}
\frac{1}{(\psi,\, \psi)}(\psi, \sum_{j=1}^n \ \mathcal{T}_j \ \psi) \geq
-\frac{16 \alpha C_1 \mathcal{L}}{3 \sqrt 3} 
\int_{\Gamma_k} B(x)^2 dx   -C_1^2n\ .
\end{equation}

By the arithmetic geometric mean inequality ($a^2 +C_1^2 \geq 2C_1\sqrt{a}$)
\begin{eqnarray}
\sum_{j=1}^{n} \left\{ T^P_j(A) -C_1|p_j+\sqrt \alpha A(x_j)|\right\}
\geq - C_1^2 n + C_1\sum_{j=1}^{n} 
\left[2 \sqrt{ T^P_j(A)} -|p_j+
\sqrt \alpha A(x_j)|\right] \\
\geq - C_1^2 n - C_1{\rm Tr} 
\left[2   \sqrt{ T^P_j(A)} -
 |p+\sqrt \alpha A| \right]_- \ .
\end{eqnarray}
Here, $[x]_-$ denotes the negative part of $x$ (which is always $\geq 0$).

Using the inequality of Birman and Solomyak \cite{BS} (see also
\cite{Loss})
\begin{eqnarray}
{\rm Tr} 
\left[2  \sqrt{T^P_j(A)} -
 |p+\sqrt \alpha A|  \right]_-
\leq {\rm Tr}  \left[4   T^P_j(A)- 
(|p+\sqrt \alpha A|)^2
\right]_-^{1/2} \\
={\rm Tr}  \left[\left(3(p+\sqrt \alpha A)^2 + 
4 \sqrt \alpha  \sigma \cdot B \right) \right]_-^{1/2} 
\ .       \label{negpart} 
\end{eqnarray}
By the Lieb-Thirring inequality (but with the added remarks in
\cite{Loss} to avoid the factor of two) (\ref{negpart}) is bounded above by
\begin{equation}
\frac{16 \alpha \mathcal{L}}{3 \sqrt 3} \int_{\Gamma_j} B(x)^2  dx \ .
\end{equation}

The bound $W_I(\underline{X})  \leq 8$ in (\ref{w}) comes from the fact that
a point $x \in \R^3$ can lie in at most 8 cubes $\Gamma_\ell$. 
\end{proof}

%%%%%%%%%%%%%%%%%%%%%%%%%%%%%%%%%%%%%%%%%%%%%%%%%%%%%%%%%%%%%%%%%%%%%%%%%%%%%%

\section{A Simpler Theorem with a Simpler Proof}\label{simpler}

In this section we show how to obtain a lower bound on the free energy
per unit volume $f$ that is correct in all respects except that it
does not vanish as $\rho \to 0$. Not only is the proof simpler but 
some of the 
constants are also better. No localization is required.
\begin{thm}[Simplified lower bound on f] 
Assume that 
$\alpha < 1/35$ and $Z\alpha^2 < (\frac{250}{320})(\frac{1}{58.5})$. 
(This means 
$Z\leq 250$ when $\alpha = 1/137$.) Then
%%(\ref{fdecompose}) holds with
\begin{align}
f(\beta, \rho, Z) &> -kT\left\{\rho_{\rm electron}\ln 2 
+\rho_{\rm electron}\, \ln (\rho_{\rm electron})
 -\rho_{\rm electron} \right\}  \nonumber \\
&\quad\quad - kT 
\rho_{\rm electron} \, \ln \left( 8\pi (kT)^3C_3^{-3}\right)
 +f_{\rm nucleus}(\beta,  \rho_{\rm nucleus})
\  ,  \nonumber \\
&\quad\quad 
 - \frac{4\pi}{3}\frac{(468)(40)}{70} \Lambda^3 
   - \left(\frac{1}{18.6\alpha}\right)^2 \rho_{\rm electron}  -0.026 \Lambda^{4}\ , 
 \label{mainbound1'}
\end{align}
where $f_{\rm nucleus}$ is the free energy per unit volume for
non-interacting 
bosonic nuclei of mass $mM$ and
\begin{equation}
C_3\alpha = \frac{1}{18.6} -\alpha^2 \max \{64.5,\, \pi Z\}
-\frac{1}{18.6}\exp\left\{\frac{-58.5}{1-58.5 \, Z\alpha^2}\right\}
\end{equation}
 \label{mainthm2}
\end{thm}
See equation (\ref{bound}) for a more general expression.

\begin{proof}
The proof is as in the proof of Theorem \ref{mainthm} except that
the electrons are not localized (but they are confined to the domain
$\Omega$ and the wave function satisfies Dirichlet boundary
conditions on $\partial \Omega$). In other words, we eliminate
$\Phi_I $ and $\sum_I$ from the equations. The localization penalty
$128 N \Lambda^{2s}$ is eliminated. 
The function $W_I(y) $ is replaced by
the characteristic function of the domain $\Omega$ and
(\ref{wprop}) is replaced by
$\int_{\R^3} W(y)dy = |\Omega|$ and $W(y) \leq 1$.

Expression (\ref{tbound2}) is  replaced, therefore, by
\begin{equation}\label{tbound22} \tag{\ref{tbound2}$'$}
-\frac{16\alpha C_1 {\mathcal{L}}}{3\sqrt3} \int_{\Omega}
B(y)^2dy -NC_1^2 \ .
\end{equation}
Thus, we save a factor of 8 because there is no longer a concern
about overlapping cells $\Gamma_\ell$. 

The function $G(y,\underline{X})$ is replaced simply by the 
characteristic function of $\Omega$ (for all $\underline{X}$), whence
$\int_{\R^3} G(y,\underline{X})dy = |\Omega|$.
The
bound (\ref{G})  is replaced by $G(y,\underline{X})\leq 1$.
In view of this, the number $\varepsilon$ in (\ref{epsilon}) 
and (\ref{eps2}) is reduced
by a factor of 8 to
\begin{equation} \label{epsilon2} \tag{\ref{epsilon}$'$}
\varepsilon' =8 \pi C_4 = 512 \pi \alpha C_1{\mathcal{L}}/3\sqrt{3} 
\end{equation}

Inequality (\ref{commutator}) becomes
\begin{equation}\label{commutator22} \tag{\ref{commutator}$'$}
\int_{\Omega} B(y)^2  dy \leq 4\int_\Omega D^*(y)\cdot  D(y) 
dy +
 \frac{1}{\pi} \Lambda^{4}\, |\Omega| \ .
\end{equation}

Lemma \ref{boxenergy} remains true, but with the obvious replacement
of 
$\sum \Phi_I^2 \int_{\R^3} B^2 W_I$ by $\int_\Omega B^2 $
in (\ref{lowerbound}) and with (\ref{w}) eliminated altogether.

Lemma \ref{gothomp} becomes

\begin{lm} \label{gothomp2}
\begin{multline}\label{lem32'}
\tr e^{-\beta H} \leq  Z_{\rm nucleus} (K) \frac{1}{N!}
\left[ \int_{\R^3} e^{-\beta C_3|p|} d p \right]^N  \ 
{\rm Tr}_{\mathcal F}\, e^{-\beta H_f} \times \\  
\exp\left\{\beta \left( + NC_1^2 + \frac{16\alpha C_1 \mathcal{L} }
{3 \sqrt 3 \pi} \Lambda^{4}|\Omega| + \frac{4 \pi}{3}|\ln(1-\varepsilon')|
\Lambda^{3}|\Omega| \right)\right\} \ ,
\end{multline} 
where $\Z_{\rm nucleus} (K)$ is the partition function of $K$
non-interacting bosons of mass $M$ (even if some of the nuclei happen
to be fermions).  The number $\varepsilon'$ (assumed to be $<1$) is in 
(\ref{epsilon2}) and, as in (\ref{see3}), 
$C_3= C_1 - \max\{64.5, \pi Z\}\alpha >0$. 
\end{lm}

The final task is to choose $C_1$. Our conditions on $\varepsilon'$ and on 
$C_3$ lead, as before, to conditions on $\alpha $ and $Z\alpha^2$, namely
$(18.6)(64.5)\alpha^2 <1$ (or $\alpha < 1/35$) and
$(18.6)\pi Z\alpha^2 <1$ (or $Z\alpha^2 < 1/58.5$). We choose a slightly lower
bound for $Z\alpha^2$, namely $Z\alpha^2 < (250/320)(1/58.5)$, and we choose 
\begin{equation}
\varepsilon' = 18.6 C_1 \alpha = 1-\exp\left(\frac{-58.5}
{1-58.5\, Z\alpha^2}\right)\ .
\end{equation}

We can bound $\frac{16 \alpha C_1 \mathcal{L}}{3\sqrt{3} \pi}$ by
$0.47/18.6 = (0.0032)8 = 0.026$. These  choices lead to Theorem 
\ref{mainbound1'}.
\end{proof}
%%%%%%%%%%%%%%%%%%%%%%%%%%%%%%%%%%%%%%%%%%%%%%%%%%%%%%%%%%%%%%%%%%%%%%%%%%%%%

\appendix \section{The Schr\"odinger representation} \label{schroe}

In the proof of Lemma \ref{gothomp}, especially eq. (\ref{separated}),
we evaluated a trace over the full Hilbert space in the
``Schr\"odinger representation" in which the field $A$ is regarded as
a c-number field.  For a fuller discussion and justification of this
method we can refer, for example, to \cite[part I, sec.
2]{glimmjaffe}, but here we discuss only what is needed in our
application.

First, we note that since the volume of the universe $|{\mathcal{U}}|$
is finite and there is an ultraviolet cutoff $\Lambda$, there are only
finitely many photon modes that interact with the electrons and
nuclei.  Each mode is a harmonic oscillator mode and can be described
in the usual Schr\"odinger representation by the canonical operators
$p_k$ and $q_k$, one pair for each $k$-value and each polarization.
In our case the $q_k$ is just the Fourier component of $A(y)$ namely
$\widehat{A}(k)$.

The noninteracting modes are infinite in number but they can be ignored
since their contribution to the trace is easy to compute (Planck's formula).

In evaluating the trace on the right side of (\ref{gt}) we can use the
$q$ representation for the photon modes and the $x$ representation for
the $L^2$ space, as usual.  The operator $H_1^{\mathrm{rel}}$
involves the electron/nuclei $p$ and $x$ operators but it involves
only the $q_k$'s and not the $p_k$'s.  On the other hand, $H_1^{\mathrm{rad}}$
involves the $p_k$ and $q_k$ operators and the $x$ operators, but it
does not involve the electron/nuclei $p$ operators.

Thus, $e^{-\beta H_1^{\mathrm{rel}}}$ is a multiplication operator as
  far as the photon modes are concerned and $e^{-\beta
    H_1{\mathrm{rad}}}$ is a multiplication operator for the
  electron/nuclei $L^2$ space.  In (\ref{separated}) the notation $d
  \underline{X} \, d\underline{R}$ is standard Lebesgue measure while
  $\mathcal{D}(A)$ means Lebesgue integration over the (finitely many)
  $q_k$'s.  It is well known that the (finite) trace of the
  exponentials of the operators that we are considering can be
  evaluated in the Schr\"odinger representation by taking the $x,x$
  and $q,q$ matrix elements and integrating over these variables in
  this manner.  Indeed, the kernel $\langle A | e^{-\beta
    H_1^{\mathrm{rad}}} | A' \rangle$ is an ($\underline{X}, \,
  \underline{R}$ dependent) Mehler kernel.

\section{A lemma about localization}

\begin{lm}[localization and convex functions] \label{loc}
Let $H$ be a self-adjoint operator on some Hilbert spaces $\mathcal{H}$
with dense domain $D(H)$. Let $\Phi_j, j=1, \dots M$ be a collection of 
bounded operators such that
$\Phi_j$ maps $D(H)$ to itself and such that
\begin{equation}
\sum_{j=1}^M \Phi_j^* \Phi_j = I    %%\sum_{j=1}^M \Phi_j \Phi_j^* = I \ .
\end{equation}
Then, for any convex function $f(x)$ and any normalized $\Psi \in D(H)$, we 
have that
\begin{equation} \label{state}
\sum_{j=1}^M \left(\Phi_j \Psi, f(H) \Phi_j \Psi \right) \geq 
f\left(\sum_{j=1}^M \left( \Phi_j \Psi, H \Phi_j \Psi\right)\right) \ .
\end{equation}
If we also assume that $\sum_{j=1}^M \Phi_j \Phi_j^* = I $
and that $f(H)$ is trace class, then
\begin{equation} \label{trace}
{\rm Tr} f(H) \geq 
{\rm Tr}f(\sum_{j=1}^M \Phi_j^* H \Phi_j)\ .
\end{equation}
\end{lm}

\begin{proof}
Note that     
\begin{multline}
\sum_{j=1}^M \left(\Phi_j \Psi,   f(H) \Phi_j \Psi\right)
=\sum_{j=1}^M \frac{\left(\Phi_j \Psi,   f(H) \Phi_j \Psi \right)}
{\left(\Phi_j \Psi, \Phi_j \Psi \right)}
\left(\Phi_j \Psi , \Phi_j \Psi \right)\\
\geq \sum_{j=1}^M f\left( \frac{\left(\Phi_j \Psi, H \Phi_j \Psi \right)}
{\left(\Phi_j \Psi, \Phi_j \Psi \right)}\right)
\left(\Phi_j \Psi, \Phi_j \Psi \right) \ , 
\end{multline}
by applying Jensen's inequality in the spectral representation of $H$.
With $c_j =  \Vert \Phi_j \Psi \Vert^2 $,  we have that 
$\sum_{j=1}^M c_j = \Vert \Psi \Vert^2=1$, and hence we may  apply Jensen's 
inequality 
once more to obtain
\begin{equation}
\sum_{j=1}^M f\left( \frac{\left(\Phi_j \Psi, H \Phi_j \Psi \right)}
{\left(\Phi_j \Psi, \Phi_j \Psi \right)}\right)
\left(\Phi_j \Psi, \Phi_j \Psi\right)
\geq f\left(\sum_{j=1}^M \left(\Phi_j \Psi, H \Phi_j \Psi\right) \right) \ .
\end{equation}
This proves (\ref{state}). To prove (\ref{trace}) we just sum (\ref{state})
over the orthonormal basis $\{\psi_n\}$  of eigenfunctions of $\sum_{j=1}^M 
\Phi_j^* H \Phi_j$ and obtain
\begin{multline}
{\rm Tr}f(\sum_{j=1}^M \Phi_j^* H \Phi_j) =\sum_n f((\psi_n, \sum_{j=1}^M 
\Phi_j^* H \Phi_j, \psi_n))
= \sum_n f(\sum_{j=1}^M (\Phi_j \psi_n, H \Phi_j \psi_n )) \\
\leq \sum_n \sum_{j=1}^M (\Phi_j \psi_n, f(H) \Phi_j \psi_n)  =
{\rm Tr}\sum_{j=1}^M
\Phi_j^* f(H) \Phi_j = {\rm Tr}\sum_{j=1}^M  f(H) \Phi_j \Phi_j^* 
={\rm Tr} f(H) \ .
\end{multline}
\end{proof}

\section{Perturbed black-body radiation}\label{radiation}

One of the problems in section \ref{strategy} is to estimate the
partition function of the Hamiltonian $H_1^{\rm rad}$ (see (\ref{rad}))
in terms of the
partition function of the universe, $\mathcal{U}$, with an error term
that depends only on the number of electrons and nuclei and, possibly,
on the ultraviolet cutoff.

In general, let us consider a Hamiltonian on Fock space of the form
\begin{equation} \label{genham}
H_M=\hbar c \sum_{k,k'} \left\{\delta_{k,k'} \delta^{i,j} |k| - M^{i,j}_{k,k'}
\right\}
a^*_i(k) a_j(k) \  ,
\end{equation}
with $M$ self adjoint, i.e.  $M^{i,j}_{k,k'}=
{\overline{M^{j,i}_{k',k}}}$.  Recall that $k=
\frac{2\pi}{L}(n_1,n_2,n_3)$ with $n_i$ an integer (but $k=0$ is
omitted) and $i,j \in \{1,2,3,\}$. The volume of $\mathcal{U}$ is $L^3$.

The   matrix $\{\cdot\}$ appearing
in (\ref{genham}),  can be written, in an obvious notation, as
$K-M$. 

Later on we shall present the $M$ we are interested in for the
purposes of this paper, but for the moment let us consider the
partition function
\begin{equation}
\ln \Z_M = \ln \tr_{\mathcal{F}} e^{-\beta H_M} =- \sum_j \ln (1-e^{-\beta \lambda_j}) 
= -\tr \ln \left\{  1- e^{-\beta (K-M)} \right\}\ ,
\end{equation}
where the $\lambda_j$ are the eigenvalues of the matrix $K-M$. In
order to make sense of $\Z_M$ we require the eigenvalues of $K-M$ to be
all positive.

We want to find an upper bound to $\ln \Z_M -\ln \Z_0$. Our main result
here is the following.  (Remark: Notice that there is no $\beta$ in
(\ref{diff}) ).

\begin{lm} \label{partfcn}
Assume that $0 \leq M < \varepsilon K$ with $0< \varepsilon <1$. Then
\begin{equation}
\ln \Z_M - \ln \Z_0 \leq \int_0^1 \tr((K-sM)^{-1} M)ds = \tr
 \ln(K) - \tr \ln(K-M) 
\leq  \left| \frac{\ln(1-\varepsilon)}{\varepsilon}\right| 
{\rm Tr}(K^{-1}M) \ . \label{diff}
\end{equation}
\end{lm}

\begin{proof}
We write the difference on the left side of (\ref{diff})
as
\begin{equation}
\int_0^1 \frac{d}{ds}\left[-\tr \ln   (1-e^{-\beta(K-sM)}) \right]  ds
= \int_0^1 \beta \tr \frac {1}{ e^{+\beta(K-sM)}-1} M ds  
\leq   \int_0^1 {\rm Tr}\frac{1}{(K-sM)} M ds \label{int}
\end{equation}
since $(e^x-1)^{-1} < x^{-1}$ for $x>0$ and since $M$ is positive
semidefinite. The estimate on the right side of (\ref{diff}) follows
by substituting $M < \varepsilon K$ in the denominator of the last
expression in (\ref{int}) and using the fact that $x^{-1}$ is matrix
monotone for $x>0$.  Finally, doing the $s$ integral we obtain the
inequality in (\ref{diff}).
\end{proof}

We now apply this lemma to the operator $H_1^{\rm rad}$
in (\ref{rad}). The matrix $M\geq 0$
is given by
\begin{equation}
M^{i,j}_{k,k'}= \frac{C_4}{(2\pi)^2} \left(\frac{2 \pi}{L}\right)^3 
\widehat{\chi}(k) \widehat{\chi}(k')
|k|^{1/2}|k'|^{1/2} \widehat G(k-k',\underline X)\left[\delta^{i,j} 
+\frac{(k\cdot k')( k_i k'_j)}{|k|^2 \, |k'|^2} - \frac{k_i'k_j'}{|k'|^2}
-\frac{k_ik_j}{|k|^2}\right] \ , 
\end{equation}
with $C_4= 64\alpha C_1 {\mathcal{L}}/3\sqrt{3}$ and where $ \widehat
G(k, \underline{X})= \int_{\R^3} e^{iy\cdot k}G(y,\underline{X}) dy$
with
$G(y, \underline X)$ given in (\ref{G}). (Here, $\underline {X} $
merely plays the role of a parameter).  First, we note that, as
matrices, $M \leq N$ where
\begin{equation}
N^{i,j}_{k,k'} = \frac{C_4}{\pi^2}
 \delta^{i,j} (\frac{2 \pi}{L})^3 \widehat{\chi}(k) 
\widehat{\chi}(k')
|k|^{1/2}|k'|^{1/2} \widehat G(k-k',\underline X) \ .
\end{equation}
The requirement that $N \leq \varepsilon K$, as a matrix, is
equivalent to the requirement that $K^{-1/2} N K^{-1/2} \leq
\varepsilon I$. Hence, we need to show that
\begin{equation}\label{c7}
\frac{C_4}{\pi^2}(\frac{2 \pi}{L})^3 \sum_{k,k'} \widehat G(k-k',\underline X) 
\widehat{\chi}(k)\overline f(k)  \widehat{\chi}(k') f(k')
\leq \varepsilon \sum_k |f(k)|^2 \ ,
\end{equation}
for all functions $f(k)$. The inequality
\begin{equation}\label{c8}
 \frac{C_4}{\pi^2}(\frac{2 \pi}{L})^3 \sum_{k,k'} \widehat G(k-k',\underline X) 
\overline f(k)   f(k')
\leq \varepsilon \sum_k |f(k)|^2 \ ,
\end{equation}
would clearly imply (\ref{c7}), and this is implied by
\begin{multline}
\frac{C_4}{\pi^2}(\frac{2 \pi}{L})^3 \int_{\R^3} G(y, {\underline{X}})
 |\sum_k e^{-iy \cdot k} f(k)|^2 dy
\\ \leq  \frac{C_4}{\pi^2} \sup_{y,{\underline{X}}} G(y,{\underline{X}})  
(\frac{2 \pi}{L})^3 \int_{\R^3}
|\sum_k e^{-iy\cdot k}  f(k)|^2 dy 
= 8\pi C_4  \sup_{y,{\underline{X}} }
G(y, \underline X) \sum_k |f(k)|^2\ .  \label{positivity}
\end{multline}

Thus, the condition 
$ 8 \pi C_4  \sup_{y, \underline{X}} G(y, \underline X) \leq \varepsilon$
guarantees that $M \leq \varepsilon K$.
If we take 
\begin{equation}\label{eps2}
\varepsilon =  64\pi C_4 = 4096\, \pi \alpha C_1{\mathcal{L}}/3\sqrt3 \ ,
\end{equation}
and use the fact that  $G(y, \underline X) \leq 8$, the condition is 
satisfied.

It remains to apply the lemma above to this particular choice of $M$, which 
(recalling (\ref{wprop})) yields the bound
\begin{align}
\ln \Z_M -\ln \Z_0 & \leq   8 \pi C_4 
\left|\frac{\ln(1-\varepsilon)}{\varepsilon}\right| 
 (\frac{2 \pi}{L})^3 
\sum_k \widehat{\chi}_\Lambda (k)^2 \int_{\R^3}G(y, \underline X)dy \\
& \approx  8 \pi C_4 
\left|\frac{\ln(1-\varepsilon)}{\varepsilon}\right| 
 \int_{\R^3} \widehat{\chi}_\Lambda (k)^2 dk \int_{\R^3} G(y, 
\underline X)dy \\
& \leq  \frac{32 \pi^2 C_4 }{3}
\left|\frac{\ln(1-\varepsilon)}{\varepsilon}\right|   
\Lambda^3 \int_{\R^3} G(y, \underline X)dy \\ 
& =\frac{4\pi}{3}|\ln(1-\varepsilon)| \Lambda^{3-3s} N
\ ,  \label{energydifference}
\end{align}
as used in (\ref{eps}).

\end{document}